\documentclass{article}
\usepackage[a4paper, margin=2.5cm]{geometry}

\usepackage{overpic}
\usepackage{natbib}

\newcommand{\Fig}[1]{Fig~\ref{fig:#1}}
\newcommand{\Eq}[1]{Eq~(\ref{eq:#1})}
\newcommand{\Dp}[2]{\frac{\partial #1}{\partial #2}}
\newcommand{\Sec}[1]{Sec~\ref{sec:#1}}
\newcommand{\dr}{\mathrm{d}}

\newcommand{\er}{\mathrm{e}}
\newcommand{\vs}{v_\mathrm{s}}
\newcommand{\kp}{k_\mathrm{p}}
\newcommand{\ka}{\overline{k}}
\newcommand{\omp}{\omega_\mathrm{p}}
\newcommand{\alp}{\alpha}
\newcommand{\erfc}{\mathrm{erfc}}

\usepackage{amsmath,amssymb}

\begin{document}

\title{Shear and depth-averaged Stokes drift under a Phillips-type spectrum}


\author{{\O}yvind Breivik\footnote{Norwegian Meteorological Institute, All\'{e}g 70, NO-5007 Bergen, Norway and the Geophysical Institute, University of Bergen.
E-mail: oyvind.breivik@met.no. ORCID Author ID: 0000-0002-2900-8458}}

\maketitle
\abstract{The transport and shear under a Phillips-type spectrum are presented. A combined profile for  monochromatic swell  and a 
Phillips-type wind sea spectrum which can be used to investigate the shear under crossing seas is then presented.}

\section{Introduction}
\label{sec:intro}
The Stokes drift profile under a Phillips-type \citep{phillips58} spectrum was explored by \citet{bre14} and 
later in more detail by \citet{bre16}. The profile was shown to give a good approximation to the profile 
under an arbitrary spectrum. Here I explore some of the features of the profile under such a spectrum. In 
\Sec{theory} I derive an expression for the depth-averaged profile. This is needed to compute the depth-
averaged Stokes drift for an arbitrary portion of the water column, which in turn is required in order to 
estimate the Stokes drift experienced by an object immersed in the water column. This has practical 
implications for the computation of the trajectories of drifting objects 
\citep{bre08,bre11,bre12,bre12b,bre13} as well as for the fate of oil in the ocean \citep{mcwilliams00}. In 
\Sec{shear} I present the general shear of the Phillips-type spectrum. This is required for the computation 
of the Langmuir-production term in the turbulent kinetic energy equation \citep{mcwilliams97}. In 
\Sec{comboprofile} I derive a combined profile for a monochromatic swell component and a wide-band 
wind sea spectrum where I assume the Phillips profile. This is of interest for investigations of the impact of 
crossing swell and wind sea on Langmuir turbulence \citep{vanroekel12,mcwilliams14,li15}.

\section{The Stokes drift transport and depth-averaged Stokes drift under the Phillips spectrum}
\label{sec:theory}
For a directional wave spectrum $E(\omega,\theta)$ the Stokes drift velocity in deep water is given by
\begin{equation}
   \mathbf{v}_\mathrm{s}(z) = \frac{2}{g} \int_0^{2\pi}\!\int_0^{\infty}
   \omega^3 \hat{\mathbf{k}} \er^{2kz}
    E(\omega,\theta) \, \dr \omega \dr\theta,
   \label{eq:uvfom}
\end{equation}
where $\theta$ is the direction in which the wave component is travelling,
$\omega$ is the circular frequency and $\hat{\mathbf{k}}$ is the unit vector in the direction of wave 
propagation.  This can be derived from the expression for a wavenumber spectrum in arbitrary depth 
first presented by \citet{kenyon69} by using the deep-water dispersion relation $\omega^2 = gk$. For 
simplicity we will now investigate the Stokes drift profile under the one-dimensional frequency 
spectrum
\begin{displaymath}
   F(\omega) \equiv \int_0^{2\pi} E(\omega,\theta) \dr \theta,
\end{displaymath}
for which the Stokes drift speed is written
\begin{equation}
   \vs(z) = \frac{2}{g}\int_0^{\infty} \omega^3F(\omega)\er^{2kz} \, \dr \omega.
   \label{eq:vs}
\end{equation}
From \Eq{vs} it is clear that at the surface the Stokes drift is proportional
to the third spectral moment [where the $n$-th spectral moment of the circular
frequency is defined as $m_n = \int_0^\infty \omega^n F(\omega) \, \dr\omega$],
\begin{equation}
   v_0 = 2m_3/g.
   \label{eq:v0}
\end{equation}

The Phillips spectrum \citep{phillips58}
\begin{equation}
   F_\mathrm{Phil} = \left\{ \begin{array}{lr}
             \alp g^2 \omega^{-5}, & \omega > \omp \\
             0,                                 & \omega \leq \omp 
                          \end{array} \right.,
   \label{eq:phil}
\end{equation}
yields a reasonable estimate of the part of the spectrum which contributes most to the Stokes drift velocity 
near the surface, i.e., the high-frequency waves. Here $\omega_\mathrm{p}$ is the peak frequency. We 
assume Phillips' parameter $\alp = 0.0083$. The Stokes drift velocity profile under (\ref{eq:phil}) is
\begin{equation}
   v_\mathrm{Phil}(z) = 2\alp g \int_{\omp}^\infty 
       \omega^{-2} \er^{2\omega^2z/g} \,\dr\omega.
   \label{eq:vphil}
\end{equation}
An analytical solution exists for (\ref{eq:vphil}), see 
\citet{bre14}, Eq~(11), which after using the deep-water dispersion relation can be written as
\begin{equation}
   v_\mathrm{Phil}(z) = \frac{2\alp g}{\omp}
   \left[\er^{-2\kp |z|} -
    \sqrt{2\pi \kp |z|}\,\erfc \left(\sqrt{2\kp |z|}\right)
     \right].
     \label{eq:vphilsolved}
\end{equation}
Here $\erfc$ is the complementary error function and $\kp = \omp^2/g$ is the peak wavenumber. From 
(\ref{eq:vphilsolved}) we see that for the Phillips spectrum (\ref{eq:vphil}) the surface Stokes drift velocity 
is $2\alp g/\omp$.

Let us assume, like \citet{bre16} do (hereafter BBJ), that the Phillips spectrum profile (\ref{eq:vphilsolved})
is  a reasonable approximation for Stokes drift velocity profiles under a general spectrum, 
\begin{equation}
v_s(z) = v_0
          \left[\er^{-2\ka|z|} - \beta \sqrt{2\pi\overline{k} |z|} \, \erfc 
              \left(\sqrt{2\ka |z|}\right) \right],
   \label{eq:vz}
\end{equation}
The total Stokes transport $V = \int_{-\infty}^0 v \, \dr z$ under \Eq{vz} can be found [see 
Appendix A of BBJ]  to be
\begin{equation}
   V = \frac{v_0}{2\ka}(1-2\beta/3).
   \label{eq:transp}
\end{equation}
We can now determine the inverse depth scale $\overline{k}$, given an estimate of the transport $V$ and 
the surface Stokes drift velocity $v_0$. Both, as \citet{bre14} argued, are normally available from wave 
models.

Note that we still need to estimate $\beta$, but BBJ found $\beta = 1$ to be a
very good approximation. Note also that estimating the Stokes transport from a
one-dimensional Stokes drift profile will overestimate it since the directional
spreading of waves tends to cancel out some of the contributions. This effect is
ignored by assuming all waves to be propagating in the same direction. The
Stokes transport should typically be reduced by about 17\% \citep{ardhuin09,
bre14}. The spreading factor proposed by \citet{webb15} could be used to further
correct the profile.

\begin{figure}[h]
\begin{center}
  \includegraphics[scale=0.6]{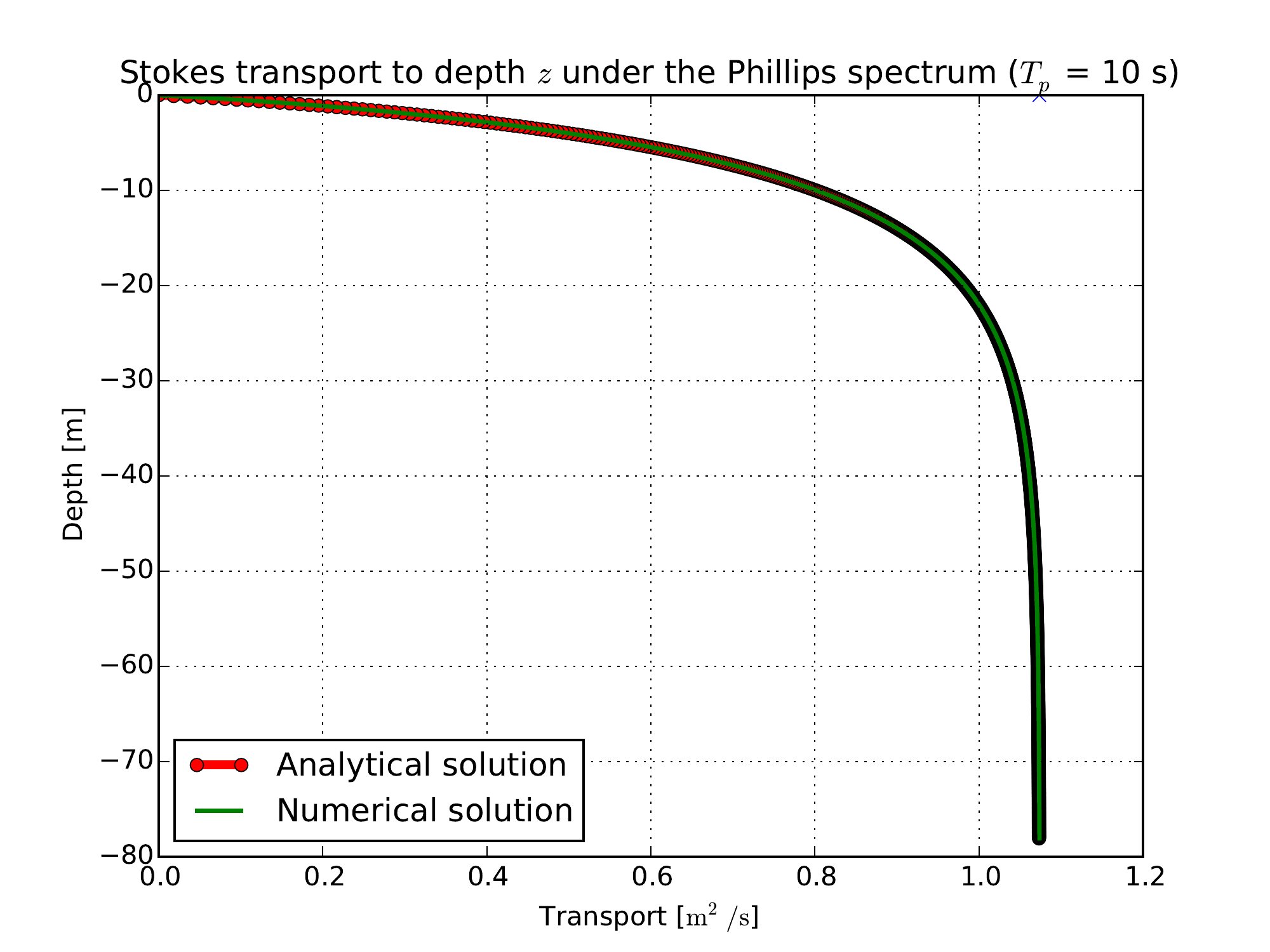}
\caption{The Stokes transport between the surface and level $z$ under a Phillips
spectrum (peak period 10 s). The analytical solution (\ref{eq:deltaV}) is
compared to a numerical integral and found to be identical to the numerical integration within numerical
accuracy.}
\label{fig:deltaVphil}
\end{center}
\end{figure}

It is useful to know the average Stokes drift over a part of the water column,
for example to compute the average Stokes drift experienced by a submerged
object. By again assuming the Phillips profile it is possible to provide a
closed-form expression for the integral of \Eq{vz}, i.e., the Stokes drift
transport between the vertical level $z_0$ and the surface,
\begin{equation}
  V(z_0) = \frac{v_0}{2\ka}\left\{1 - \er^{-2\ka |z_0|} - \frac{2\beta}{3}\left[1 + \sqrt{\pi}(2\ka |z_0|)^{3/2} \,
      \erfc\,\left(\sqrt{2\ka |z_0|}\right) 
                  -(1+2\ka |z_0|) \er^{-2\ka |z_0|}\right]\right\}.
     \label{eq:deltaV}
\end{equation}
See the appendix for a full derivation. 
Obviously, in order to find the average Stokes drift between a lower level $z_0$ and an upper level $z_1$ all that is needed is to use 
\Eq{deltaV} twice to find 
\begin{equation}
   \overline{v}_\mathrm{s} = \frac{V(z_0)-V(z_1)}{z_1-z_0}.
   \label{eq:vavg}
\end{equation}
\Fig{deltaVphil} shows the transport under the Phillips spectrum (10 s peak period) integrated to depth $z_0$. The analytical solution (\ref{eq:deltaV}) is found to be identical to the numerical integration within numerical accuracy. 
\Eq{deltaV} represents the transport between level $z_0$ and the surface. 

\section{The shear under the Phillips spectrum}
\label{sec:shear}
The shear under \Eq{vz} is straightforward to find,
        \begin{equation}
             \Dp{\vs}{z} = v_0\left[2(1-\beta)\ka \er^{-2\ka|z|} + \beta \sqrt{\frac{\pi \ka}{2 |z|}} \, \erfc  
               \left(\sqrt{2\overline{k}|z|}\right)\right],   
               \label{eq:shear}  
        \end{equation}
which simplifies to
        \begin{equation}
             \Dp{\vs}{z} = v_0\sqrt{\frac{\pi \overline{k}}{2 |z|}} \, \erfc  
               \left(\sqrt{2\overline{k}|z|}\right)   
               \label{eq:shearbeta1}  
        \end{equation}
when $\beta = 1$.

\section{A combined Stokes profile for swell and wind sea}
\label{sec:comboprofile}
Let us decompose the Stokes drift velocity in a swell component and a wind sea component,
\begin{equation}
   \mathbf{v}_\mathrm{s} = \mathbf{v}_\mathrm{sw} + \mathbf{v}_\mathrm{ws} .
     \label{eq:vswws}
\end{equation}
Let us assume that the swell is well represented by the monochromatic profile such that  
\begin{equation}
   \mathbf{v}_\mathrm{sw} = v_\mathrm{sw,0} \er^{2k_\mathrm{sw} z} \hat{\mathbf{k}}_\mathrm{sw}.
     \label{eq:vsw}
\end{equation}
Here $\hat{\mathbf{k}}_\mathrm{sw}$ is a unit vector in the swell propagation
direction $\theta_\mathrm{sw}$.  The wavenumber is found from the linear
deep-water dispersion relation to be $k_\mathrm{sw} = 4\pi^2f^2_\mathrm{sw}/g$.
The surface swell Stokes drift speed is
\begin{equation}
   v_\mathrm{sw,0} = 2k_\mathrm{sw}V_\mathrm{sw}.
   \label{eq:vsw0}
\end{equation}
The swell Stokes transport can be found from the swell height and frequency \citep{bre14},
\begin{equation}
   V_\mathrm{sw} = 2\pi m^\mathrm{sw}_1 = \frac{2\pi}{16}f^\mathrm{sw}_{m01}H_\mathrm{sw}^2,
   \label{eq:Vsw}
\end{equation}
from which the swell wavenumber can be computed,
\begin{equation}
   k_\mathrm{sw} = v_\mathrm{sw,0}/2V_\mathrm{sw}.
   \label{eq:ksw}
\end{equation}

Assume that \Eq{vz} is a good approximation for the wind sea part of the Stokes
profile, and let the surface Stokes drift from the wind sea part of the spectrum
be defined as
\begin{equation}
   v_\mathrm{ws,0} = |\mathbf{v}_\mathrm{s,0} - \mathbf{v}_\mathrm{sw,0}| .
     \label{eq:vws0}
\end{equation}
The wind sea transport $V_\mathrm{ws}$ is determined similarly as the swell
transport (\ref{eq:Vsw}),
\begin{equation}
   V_\mathrm{ws} = 2\pi m^\mathrm{ws}_1 = \frac{2\pi}{16}f^\mathrm{ws}_{m01}H_\mathrm{ws}^2,
   \label{eq:Vws}
\end{equation}
Finally, the inverse depth scale (or Phillips peak
wavenumber) of the wind sea profile is found from \Eq{transp},
\begin{equation}
   k_\mathrm{ws} = \frac{v_\mathrm{ws,0}}{2V_\mathrm{ws}}(1-2\beta/3).
   \label{eq:kws}
\end{equation}
To ensure that the surface Stokes drift vector is preserved, the direction of the wind sea profile should be 
determined from \Eq{vswws}, $\hat{\mathbf{k}}_\mathrm{ws} = (\mathbf{v}_\mathrm{s,0} - \mathbf{v}_\mathrm{ws,0})/
v_\mathrm{ws,0}$. Note that the transport under the combined profile will be smaller or equal to the 
transport under the one-dimensional profile resulting from the total sea state parameters since the 
swell and wind sea components will tend to cancel each other out unless they are in perfect alignment 
(see the discussion by \citealt{bre14}) .

This simple procedure allows us to estimate a combined profile with a directional veering
due to the presence of swell. The parameters can all be estimated from standard
output from atmosphere-wave reanalyses such as ERA-Interim \citep{dee11} or 
regional wave hindcasts like NORA10 \citep{rei11, semedo15}.

\clearpage
\setcounter{figure}{0}

\appendix

\section{The transport under a Phillips-type spectrum}
\label{sec:philapp}
The Stokes transport under \Eq{vz} can be written [following \Eq{deltaV}]
\begin{equation}
   V(y_0) = \frac{v_0}{2\ka} \int_{y_0}^0 \! \er^{-y} \, \dr y- 
         \beta \sqrt{\pi}\underbrace{\int_{y_0}^0 \sqrt y \, \erfc \left(\sqrt{y}\right)}_{I_1} \, \dr y.
     \label{eq:deltaV2}
\end{equation}
Here we have introduced the variable substitution $y = -2\ka z$. The integral $I_1$
can be solved by first substituting $u = \sqrt{y}$,
\begin{equation}
   I_1 = 2\int_0^{u_0} u^2 \, \erfc \, u\, \dr u.
     \label{eq:substu}
\end{equation}
Integrate by parts to get
\begin{equation}
   I_1 = \frac{2}{3}u_0^3 \, \erfc \, u_0 + \frac{4}{3\sqrt \pi} \underbrace{\int_0^{u_0} u^3 \er^{-u^2} \, \dr u}_{I_2}.
     \label{eq:byparts}
\end{equation}
The integral $I_2$ can be found analytically [see Eq (3.321.6) of \citealt{gradshteyn07}],
\begin{equation}
   I_2 = 1 - (1+u_1^2) \er^{-u_1^2}.
   \label{eq:gr3.321.6}
\end{equation}
\Eq{deltaV2} can now be solved,
 \begin{equation}
   V(y_0) = \frac{v_0}{2 \ka} \left\{1 -  \er^{-y_0} - \frac{2\beta}{3}\left[1+\sqrt{\pi}y_0^{3/2} \, \erfc \, \sqrt{y_0} -(1+y_0)\er^{-y_0}\right]\right\}.
      \label{eq:deltaVy}
\end{equation}

\clearpage

\bibliographystyle{elsarticle-harv}


\end{document}